\documentclass[lettersize,journal]{IEEEtran}
\usepackage{amsmath,amsfonts}
\usepackage{algorithmic}
\usepackage{algorithm}
\usepackage{array}
\usepackage[caption=false,font=normalsize,labelfont=sf,textfont=sf]{subfig}
\usepackage{textcomp}
\usepackage{stfloats}
\usepackage{url}
\usepackage{verbatim}
\usepackage{graphicx}
\usepackage{cite}
\usepackage{booktabs}
\usepackage{multirow}
\begin{document}

\title{DGSD: Dynamical Graph Self-Distillation for EEG-Based Auditory Spatial Attention Detection}

\author{
Cunhang Fan,~\IEEEmembership{Member,~IEEE,}
Hongyu Zhang,
Wei Huang,
Jun Xue,
Jianhua Tao,~\IEEEmembership{Senior Member,~IEEE,}
Jiangyan Yi,~\IEEEmembership{Member,~IEEE,}
Zhao Lv,~\IEEEmembership{Member,~IEEE,}
Xiaopei Wu,~\IEEEmembership{Member,~IEEE}

\thanks{This work is supported by the {STI 2030—Major Projects (No. 2021ZD0201500)}, the National Natural Science Foundation of China (NSFC) (No.62201002, No.61972437), Distinguished Youth Foundation of Anhui Scientific Committee (No. 2208085J05), Special Fund for Key Program of Science and Technology of Anhui Province (No. 202203a07020008), Open Fund of Key Laboratory of Flight Techniques and Flight Safety, CACC (No, FZ2022KF15), the Open Research Projects of Zhejiang Lab (NO. 2021KH0AB06) and the Open Projects Program of National Laboratory of Pattern Recognition (NO. 202200014). (Corresponding author: Xiaopei Wu and Zhao Lv.)}
\thanks{Cunhang Fan, Hongyu Zhang, Wei Huang, Jun Xue, Zhao Lv and Xiaopei Wu are with the Anhui Province Key Laboratory of Multimodal Cognitive Computation, School of Computer Science and Technology, Anhui University, Hefei 230601, China (e-mail: cunhang.fan@ahu.edu.cn; e22201103@stu.ahu.edu.cn; 09057@ahu.edu.cn; e21201068@stu.ahu.edu.cn; kjlz@ahu.edu.cn; 88045@ahu.edu.cn).}

\thanks{Jiangyan Yi is with the National Laboratory of Pattern Recognition, Institute of Automation, Chinese Academy of Sciences, Beijing 100190, China (e-mail: jiangyan.yi@nlpr.ia.ac.cn).}

\thanks{Jianhua Tao is with Department of Automation, Tsinghua University, Beijing 100190, China (e-mail: jhtao@tsinghua.edu.cn)}}
\markboth{Journal of \LaTeX\ Class Files,~Vol.~14, No.~8, August~2021}%
{Shell \MakeLowercase{\textit{et al.}}: A Sample Article Using IEEEtran.cls for IEEE Journals}


\maketitle

\begin{abstract}
Auditory Attention Detection (AAD) aims to detect target speaker from brain signals in a multi-speaker environment. Although EEG-based AAD methods have shown promising results in recent years, current approaches primarily rely on traditional convolutional neural network designed for processing Euclidean data like images. This makes it challenging to handle EEG signals, which possess non-Euclidean characteristics. In order to address this problem, this paper proposes a dynamical graph self-distillation (DGSD) approach for AAD, which does not require speech stimuli as input. Specifically, to effectively represent the non-Euclidean properties of EEG signals, dynamical graph convolutional networks are applied to represent the graph structure of EEG signals, which can also extract crucial features related to auditory spatial attention in EEG signals. In addition, to further improve AAD detection performance, self-distillation, consisting of feature distillation and hierarchical distillation strategies at each layer, is integrated. These strategies leverage features and classification results from the deepest network layers to guide the learning of shallow layers. Our experiments are conducted on two publicly available datasets, KUL and DTU. Under a 1-second time window, we achieve results of 90.0\% and 79.6\% accuracy on KUL and DTU, respectively. We compare our DGSD method with competitive baselines, and the experimental results indicate that the detection performance of our proposed DGSD method is not only superior to the best reproducible baseline but also significantly reduces the number of trainable parameters by approximately 100 times.
\end{abstract}

\begin{IEEEkeywords}
Auditory attention detection, electroencephalography (EEG), dynamical graph convolutional network, self-distillation.
\end{IEEEkeywords}

\section{Introduction}
The cocktail party problem \cite{ref1, ref2} is an intriguing scenario where multiple speakers' voices are mixed together, much like in a noisy social gathering. The challenge in this problem arises when multiple sound sources are present simultaneously, and we need to find a way to separate and extract the sound source of interest, namely, the target speaker. Multi-speaker speech separation techniques \cite{reftaslp1, reffan} are used to address the aforementioned issue. These techniques aim to decompose mixed speech into different sound sources, allowing us to individually extract the speech of each speaker. However, these techniques cannot extract the target speech without the prior information of the target speaker. To tackle this problem, auditory attention detection (AAD) \cite{ref3, ref4, ref5, ref6} has emerged as a highly promising solution. AAD is designed to emulate the "attention" process in the human auditory system using brain signals. With AAD technology, we can identify and locate the target speaker, i.e., the speaker who has captured the listener's attention in a multi-speaker environment. Hearing-impaired individuals often struggle to differentiate between target and interfering sounds in noisy environments due to their hearing disabilities, leading to communication difficulties and emotional challenges. Modern hearing aids \cite{ref7} incorporate advanced AAD algorithms to assist hearing impaired people in more accurately capturing target sounds, ignoring background noise, and enhancing their listening performance in multi-speaker scenarios.

Electroencephalography (EEG) provides a non-invasive and low-cost technique. Various studies indicate that using EEG for AAD is feasible \cite{ref8, ref9, ref10, ref11, ref12, ref13}. AAD relies on extracting EEG features from the EEG signals, which can be done in the time domain \cite{ref14, ref15, ref16} and frequency domain. Extracting EEG features from the frequency domain allows for a more comprehensive reflection of signal characteristics compared to the time domain. The extraction of frequency domain features is used to identify different frequency bands of brainwave rhythms, such as $\delta$ (1-3 Hz), $\theta$ (4-7 Hz), $\alpha$ (8-13 Hz), $\beta$ (14-30 Hz), and $\gamma$ (31-50 Hz) \cite{ref17, ref18, ref19, ref20, ref21}. This helps describe the spatial characteristics and functional states of the EEG signals. Subsequently, EEG features can be extracted from each frequency band, including power spectral density (PSD) \cite{ref22, reftaslp2} features, rational asymmetry (RASM) \cite{ref49} features, differential entropy (DE) \cite{ref23, ref24} features, and so on.

Research on AAD primarily focuses on two paradigms \cite{ref7}: speaker identification and tracking spatial attention. The former requires both EEG signals and clean auditory stimuli as input \cite{ref25, ref26}, while the latter only relies on EEG signals \cite{ref13, ref27}. In this paper, we focus on models that use only EEG signals as input, choosing not to use auditory stimuli for practical reasons. Previous studies \cite{ref8, ref5} often use clean stimulus as input, but in real-world environments, listeners typically receive mixed speech containing voices from multiple speakers. This makes obtaining clean auditory stimuli challenging, as models might face various challenges when dealing with mixed speech, such as speech separation. Therefore, considering the complexity and feasibility of real-world applications, we choose not to use this approach to ensure the practicality and effectiveness of our model in real-world scenarios.

In recent years, due to findings in neuroscience suggesting that the brain processes auditory stimuli through nonlinear mappings \cite{ref28}, traditional linear AAD methods struggle to handle the nonlinear mappings in the brain, and their decoding performance deteriorates significantly with shorter time windows \cite{ref29}. Consequently, research has gradually shifted towards nonlinear methods based on EEG \cite{ref7}, with convolutional neural networks (CNNs) \cite{ref13, ref27, ref30} being the most commonly used nonlinear approach. When performing auditory spatial attention detection, selecting an appropriate method to model EEG signals is of paramount importance. Unlike Euclidean data such as image pixels, EEG signals exhibit non-uniform sampling due to the uneven measurement locations on the scalp and varying distances between electrodes. Additionally, during the data collection process, electrodes are distributed discretely on the scalp, forming a discrete electrode network rather than a continuous Euclidean space. The primary reason why CNN is unsuitable for processing EEG signals is that CNN is designed to handle Euclidean data, relying on spatial relationships between pixels during the convolutional kernel sliding process. However, the non-uniform sampling points and non-Euclidean spatial characteristics of EEG signals make it challenging to model such spatial relationships effectively \cite{refg1}. Relatively speaking, graph structures can naturally represent these non-uniform connections, allowing for a better capture of interactions between different electrodes. They are not constrained by a fixed grid structure and are better suited to adapt to the characteristics of EEG signals, thus providing more accurate feature extraction \cite{refg2}.

In this paper, a novel dynamical graph self-distillation (DGSD) method is proposed for auditory spatial attention detection. Initially, dynamical graph convolutional networks (DGCN) are employed to represent EEG signals with non-Euclidean features as a graph structure, where each node corresponds to an electrode location, and the adjacency matrix represents the connectivity between electrodes. Subsequently, graph convolution operations within the network are utilized to extract essential features related to auditory spatial attention. These operations propagate information between electrodes and dynamically update the feature representation of electrodes using information from neighboring electrodes. Additionally, self-distillation methods are integrated, applying feature distillation and hierarchical distillation strategies after each layer of graph convolution operations. This involves using features and classification results from the deepest layer to guide the learning of shallower layers, further enhancing the model's performance.

The main contributions of this paper lie in two aspects. Firstly, the DGCN is applied to represent EEG signals with non-Euclidean characteristics and capture crucial features related to auditory spatial attention. Secondly, the self-distillation is integrated to further improve the model's performance, which consists of feature distillation and hierarchical distillation strategies at each layer. Experiments are conducted on publicly available datasets from KUL and DTU. The experimental results demonstrate that the proposed DGSD approach not only outperforms state-of-the-art reproducible AAD methods but also significantly reduces the number of trainable parameters by approximately 100 times.

The rest of this paper is organized as follows. Section~\ref{related} presents a brief overview of the relevant work related to this paper. Section~\ref{method} introduces the proposed DGSD method. The experimental setup is stated in Section~\ref{experiment}. Section~\ref{result} shows experimental results. Section~\ref{discussion} shows the discussions. Section~\ref{conclusion} draws conclusions.

\section{Related Work}
\label{related}
\subsection{Nonlinear methods for AAD}
Currently, advanced AAD research methods can be categorized into two types. One type is related to spatial localization detection, such as \cite{ref13, ref27, refcaig}, both of which use EEG signals as input and employ CNN for spatial feature extraction. In \cite{ref27}, the authors propose the SSF-CNN method for auditory spatial attention detection, which combines spectral spatial features (SSF) constructed by analyzing the topographical specificity of $\alpha$ band power in EEG with CNN to enhance detection performance. In \cite{ref13}, the authors found that using SSF-CNN with only the $\alpha$ band power spectrum couldn't fully reflect the spatial information of EEG, so they extracted multi-band differential entropy features as input to CNN, utilizing information from different frequency bands in EEG signals to improve performance. In \cite{refcaig}, the authors propose an EEG-graphs convolutional network that incorporates a neural attention mechanism. It takes EEG data from a single $\beta$ frequency band as input, and this mechanism simulates the topological structure of the human brain based on the spatial patterns of EEG signals. The other type is speaker identification, as in \cite{ref38, ref39, ref40}, which uses both EEG signals and auditory stimuli as input. In \cite{ref38, ref39, ref40}, the authors introduce a joint CNN-LSTM model, which takes EEG signals and stimulus spectrograms as inputs for identity recognition, improving performance by capturing long-term dependencies between EEG responses and auditory stimuli using long short-term memory (LSTM). 

\subsection{Spectral graph filtering} 
Spectral filtering \cite{ref33}, also known as graph convolution, is a widely used signal processing technique in graph data operations. The basic idea is to represent graph data as a signal in a specific domain, such as the frequency domain or spectral domain. In these domains, the graph fourier transform (GFT) \cite{ref41} can be used for graph signal analysis. Recently, spectral filtering has been widely applied in graph neural network (GNN) to form graph convolutional network (GCN) \cite{ref31, ref32, ref34, ref42, ref43}, which can extract features of nodes and edges as a convolution operation. In \cite{ref44, ref48}, the authors first propose a domain-based hierarchical clustering or graph Laplacian spectrum GCN, which can handle signals on irregular graph structures such as social networks and brain connectomes. In \cite{ref21}, the authors introduce a dynamical graph convolutional neural networks (DGCNN) approach that incorporates GCN into an emotion recognition system based on multi-channel EEG. This method dynamically learns the intrinsic relationships between different EEG channels. This indicates GCN have great potential in extracting features of discrete spatial domain signals \cite{ref31}.

\subsection{Self-distillation} 
Self-distillation, as an emerging method, has been applied in various fields such as speech recognition and computer vision \cite{ref35, ref36}. In \cite{ref45}, the proposed instance segmentation network is trained and its detection accuracy is improved by applying self-distillation. In \cite{ref46}, an elegant self-distillation mechanism is proposed to directly obtain high-precision models. In \cite{ref37}, a self-distillation method for fake speech detection is proposed, which uses the deepest network to guide and enhance the shallow network, and builds a distillation path between the features of the deepest and shallow networks to reduce feature differences. This method can significantly improve the performance of FSD. These methods demonstrate that self-distillation, as an effective knowledge transfer and model training method, has broad application prospects in different fields and tasks.

\begin{figure*}[!t]
	\centering
        \includegraphics[width=2\columnwidth]{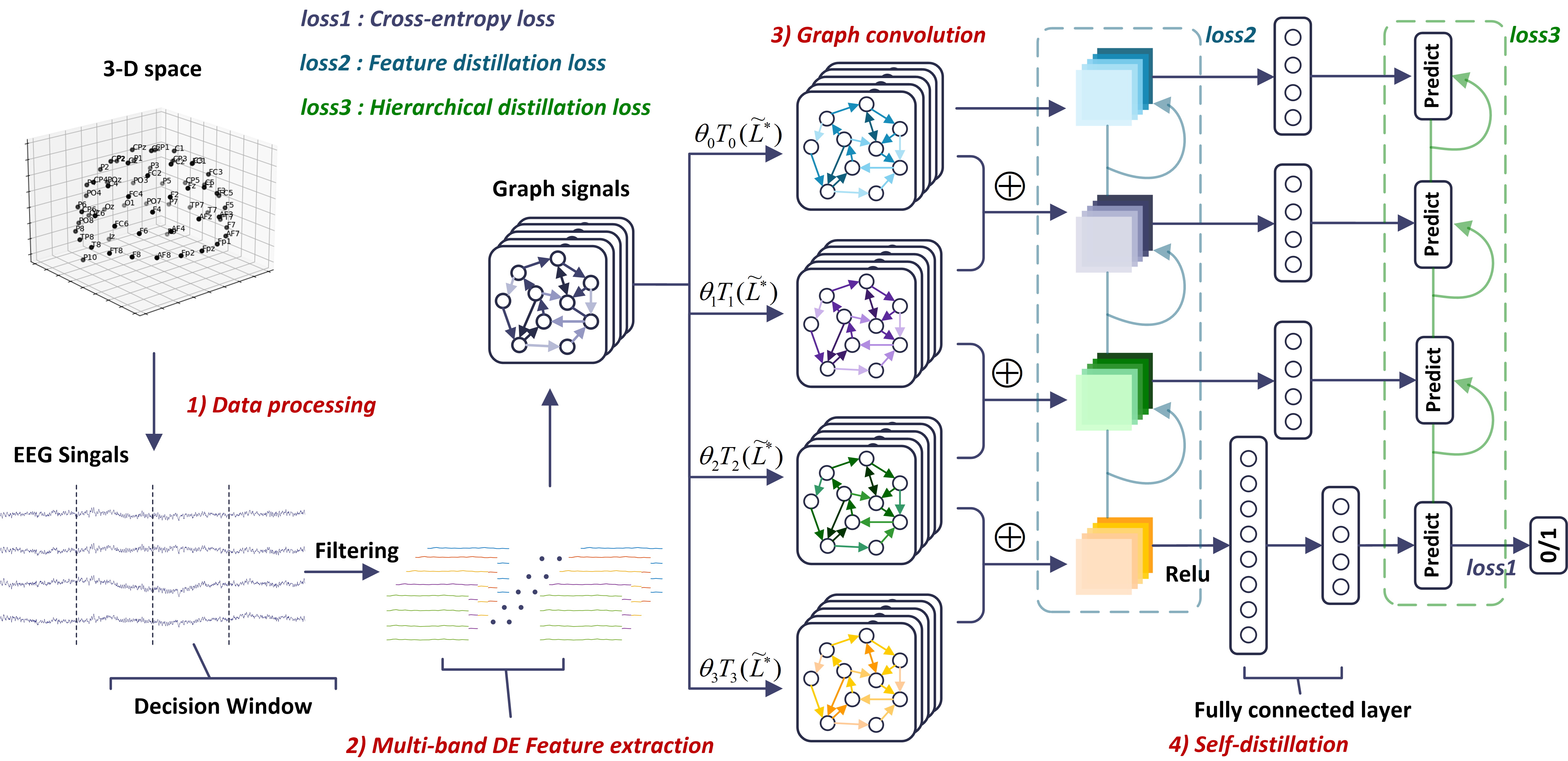}
	\captionsetup{labelfont=bf}
	\caption{DGSD model architecture. DGSD consists of four modules :1) EEG data processing, 2) Multi-band DE feature extraction, 3) Graph convolution operation and 4) Self-distillation. The fusion of the latter two modules is the core of our model. The graph structure is used to represent EEG signals, and then the graph convolution of each layer is used to extract features about auditory spatial attention from EEG signals, while the features and classification results of the deepest network are used to guide the learning of the shallow network.}
	\label{model}
\end{figure*}

\section{The proposed DGSD method}
\label{method}
In this section, we introduce our proposed DGSD model. This method not only effectively represents EEG data with non-Euclidean characteristics as graph signals but also extracts crucial features related to auditory spatial attention using graph convolution operations. Furthermore, by combining with self-distillation, the model can enhance detection accuracy, which enables the features and classification results from the deepest network layer to guide the shallow network learning through feature distillation and hierarchical distillation in self-distillation. The framework of our proposed model is illustrated in Fig.~\ref{model}, which consists of four modules: EEG data processing, multi-band DE feature extraction, graph convolution operation and self-distillation. Next, we provide detailed descriptions of these modules.

\subsection{EEG data processing \& Multi-band DE feature extraction}
\label{pro}

Many studies use the sliding window method to segment EEG signals into a series of time periods for performance analysis of different AAD algorithms \cite{ref13, ref27, ref29, ref30, ref38}. In this study, we process the data according to the final frequency of EEG preprocessing for each dataset, and perform sliding window processing for each subject's data to extract multi-band DE features in each time segment.

Next, we perform frequency band decomposition on the EEG data after sliding window. It is decomposed into five frequency bands, allowing for a comprehensive description of the spatial characteristics and functional states of the EEG signal. We then extract multi-band DE features \cite{ref23, ref50} from each frequency band. As a result, for the input EEG signal with 64 channels, we obtain a total of 320 DE features, consisting of 64 channels across the five frequency bands.

\subsection{Dynamical graph convolution network}
\label{gc}
The dynamical graph convolutional network (DGCN) represents EEG signals with non-Euclidean properties as a graph structure for subsequent graph convolution operations. Then, it can utilize graph convolution operations to extract important features related to auditory spatial attention. This network can obtain more discriminative features by dynamically updating the adjacency matrix within the graph structure. Further information about the dynamic updating of the adjacency matrix can be found in Section~\ref{ww}.

\subsubsection{\textbf{Graph representation of EEG signals}} 
At this point, we can construct a graph $ \mathbf{G} $ using the outputs of the multi-band differential entropy extraction module, which serves as the input for the graph distillation module. In $ \mathbf{G = \{V, W\}} $, $ \mathbf{V = \{ v_{1}, v_{2},\ldots, v_{N}} \} $ is the node set where each node $ \mathbf{v_{i}} $ corresponds to an electrode and $ \mathbf{N} $ is the number of electrodes in the EEG recording equipment. $ \mathbf{W \in \mathbb{R}^{N \times N}} $ is the adjacency matrix of $ \mathbf{G} $, with non-negative elements $ \mathbf{w_{ij}} $ representing the strength of functional connection between $ \mathbf{v_{i}} $ and $ \mathbf{v_{j}} $. Each node is associated with $ \mathbf{d} $ features, i.e., the feature matrix $ \mathbf{x \in \mathbb{R}^{N \times d}} $ of the nodes. Each column of $ \mathbf{x} $ represents a signal defined on the node. Our next step is to perform operations on $ \mathbf{x} $.

\subsubsection{\textbf{Graph convolution}} 
Specifically, the Laplacian matrix of graph $ \mathbf{G = \{V, W\}} $ is $ \mathbf{L = D - W}$ (where $ \mathbf{D} $ is a diagonal matrix with elements $ \mathbf{D_{ii} = \sum_{j=1}^{N}w_{ij}}$), its eigenvector matrix is $\mathbf{U = \left[ u_{1}, u_{2},\ldots, u_{N}\right]} $ and eigenvalue matrix is \(\mathbf{\Lambda = \text{diag}(\left[ \lambda_{1}, \lambda_{2}, \ldots, \lambda_{N}\right])}\), which can be obtained through the singular value decomposition of $ \mathbf{L} $, i.e., $ \mathbf{L = U \Lambda U^T} $. Then the Fourier transform of $ \mathbf{x} $ in the graph domain can be expressed as $ \mathbf{\hat{x} = U^T x} $. The graph convolution operator is defined in the graph domain as:
\begin{equation}
\label{xy}
\mathbf{x \ast y = U \left[(U^T x) \odot (U^T y)\right]}
\end{equation}
where $ \odot $ denotes the element-wise Hadamard product.

The key to graph convolution (also known as spectral filtering) is how to choose the filter $ \mathbf{g}$ to adjust the Fourier coefficients of the signal $ \mathbf{ x } $ in the spectral domain, and thus control the response of the signal at different frequencies. Typically, the filter $ \mathbf{g} $ is a diagonal matrix whose diagonal elements represent the weights at different frequencies, i.e., $ \mathbf{g(\Lambda) = \text{diag}(\left[ \theta_{1}, \theta_{2}, \ldots, \theta_{N}\right])} $, where $ \mathbf{\{ \theta_{i} \}_{i=1}^{N}} $ is the vector of Fourier coefficients. Therefore, for a signal $ \mathbf{ x } $ that has been processed by the filter $ \mathbf{g(L) }$, its Fourier coefficients can be represented as:
\begin{equation}
\label{y1}
\mathbf{ y = g(L)x = g(U \Lambda U^T)x =Ug(\Lambda)U^T x}
\end{equation}

This process can be seen as a convolution operation in the spectral domain, that is:
\begin{equation}
\label{y2}
\begin{aligned}
\mathbf{ y } & = \mathbf{Ug(\Lambda)U^T x} = \mathbf{\left[Ug(\Lambda)\right] \odot (U^T x) }\\
& \mathbf{= U\{U^T\left[Ug(\Lambda)\right]\}\odot(U^T x) = x \ast \left[ Ug(\Lambda)\right] }
\end{aligned}
\end{equation}
where $ \mathbf{Ug(\Lambda)} $ is the convolution kernel, called graph convolution operator. In this way, the signal $ \mathbf{x} $ can be transformed from the graph spatial domain to the graph spectral domain, and then convolved with the graph convolution operator, to obtain the signal $ \mathbf{y} $ processed by the filter $ \mathbf{g} $.

We adopt a similar graph convolution representation as in \cite{ref21}, which is the graph convolution modified by K-order Chebyshev polynomials. The convolution kernel formula is as follows:

\begin{equation}
\label{g_l}
\mathbf{g(\Lambda) \approx \sum_{k=0}^{K-1}\theta_{k}T_{k}(\tilde{\Lambda})} 
\end{equation}
where $ \mathbf{\theta_{k}} $ is the coefficient of the Chebyshev polynomial, $ \mathbf{\tilde{\Lambda}} $ is normalized by $ \mathbf{\Lambda} $. $ \mathbf{T_{k}(\tilde{\Lambda})} $ is the K-order Chebyshev polynomial used for evaluating $ \mathbf{\tilde{\Lambda}} $.

Currently, our graph convolution is capable of extracting features related to auditory spatial attention. In this module, we design a 4-layer graph convolution, and each layer of signal $ \mathbf{x} $ undergoes graph convolution as $ \mathbf{x_{i} = x_{i-1} + DGCN(x, A_{i})} $, where $ \mathbf{A} $ is an $ \mathbf{\tilde{\Lambda}} $ matrix generated by K-order Chebyshev polynomial. Next, we provide a detailed explanation of our self-distillation strategy incorporated in the graph convolution layers and the dynamic updating process of the adjacency matrix.

\subsection{Self-distillation}
\label{sd}
To further enhance the AAD detection performance, we incorporate the self-distillation method, which consists of feature distillation and hierarchical distillation after each DGCN layer. It guides the learning of shallow networks using features and classification results extracted by the deepest network, thereby extracting more suitable classification features for the AAD task.

\subsubsection{\textbf{The calculation of the loss composition}} 
We use the cross-entropy loss function to calculate the loss between the convolution result of the four-layer graph and the true label. This loss is the main loss function for classification, expressed as $ \mathbf{loss1} $:
\begin{equation}
\label{loss1}
\mathbf{loss1 = CrossEntropy(p_{n},y)}
\end{equation}
where $ \mathbf{p_{n}} $ is the output of the deepest DGCN, with $ \mathbf{n} $ being set to 4 in this paper. $ \mathbf{y} $ is the label of the training dataset. The process of applying average pooling after each graph convolution layer to extract task-related essential features can be described as follows:
\begin{equation}
\label{fi}
\mathbf{F_{i} = avgpool(conv(x_{i}))}
\end{equation}

Utilizing the features extracted from the shallow and deepest layers of DGCN, a feature distillation loss is generated using the L2 function. This loss encourages the shallow-layer features to adapt to the deepest layer features, while using the deepest layer features to guide the learning of shallow-layer features. As shown in Fig.~\ref{model}, this results in $ \mathbf{loss2} $. In this way, when predicting classification results, the shallow DGCN can better align with the outcomes of the deepest DGCN. The calculation formula for $\mathbf{loss2}$ is as follows:
\begin{equation}
\label{loss2}
\mathbf{loss2 = \sum_{i=1}^{n-1}L2(F_{i}, F_{n})}
\end{equation}
where $ \mathbf{F_{i}} $ is the output feature of each shallow DGCN, and $ \mathbf{F_{n}} $ is the output feature of the deepest DGCN. $ \mathbf{L2} $ is the L2 loss function. In addition, we design a classifier for $ \mathbf{x} $ after each layer of graph convolution, which generates $ \mathbf{M} $ classification results for subsequent guidance of the deepest DGCN on the shallow DGCN in a global sense. It is worth noting that these classifiers are only used for training and not used in validation and testing phases. We take the deepest DGCN (i.e., the fourth layer) as the teacher model and the first three DGCNs as the student model. Then we use KL divergence to calculate the hierarchical distillation loss in the teacher-student model, i.e., $ \mathbf{loss3} $ in Fig.~\ref{model}, which can obtain the difference between the two output distributions and better guide the shallow network in learning features. The calculation formula of $ \mathbf{loss3} $ is as follows:
\begin{equation}
\label{loss3}
\mathbf{loss3 = \sum_{i=1}^{n-1}KL(p_{i}, p_{n})}
\end{equation}
where $ \mathbf{p_{i}} $ represents the output of each layer in the network after the fully connected classifier, and $ \mathbf{KL} $ refers to the KL divergence function. At this point, the training loss consists of three components, where $\alpha$ and $\beta$ are hyperparameters that balance these three sources of loss. Both hyperparameters have values between 0 and 1. The final $ \mathbf{loss} $ is:
\begin{equation}
\label{loss}
\mathbf{loss = \alpha loss1 + (1-\alpha) loss2 + \beta loss3}
\end{equation}

\subsubsection{\textbf{Dynamic learning of the adjacency matrix $ \mathbf{W} $}} 
\label{ww}
We use the back-propagation (BP) method to iteratively update network parameters during model training to achieve optimal or suboptimal solutions. To dynamically learn the optimal adjacency matrix $ \mathbf{W} $ of the DGSD model using the BP method, we must compute the partial derivative of the loss function with respect to $ \mathbf{W} $, which is expressed as follows:
\begin{equation}
\label{loss_w}
\mathbf{\frac{\partial loss}{\partial W} = 
\begin{pmatrix}
	\mathbf{\frac{\partial loss}{\partial w_{11}}} & \mathbf{\frac{\partial loss}{\partial w_{12}}} & \mathbf{\cdots} & \mathbf{\frac{\partial loss}{\partial w_{1N}}} \\
	\mathbf{\vdots} & \mathbf{\vdots} & \mathbf{\frac{\partial loss}{\partial w_{ij}}} & \mathbf{\vdots} \\
	\mathbf{\frac{\partial loss}{\partial w_{N1}}} & \mathbf{\frac{\partial loss}{\partial w_{N2}}} & \mathbf{\cdots} & \mathbf{\frac{\partial loss}{\partial w_{NN}}} \\
\end{pmatrix}}
\end{equation}
where $ \mathbf{w_{ij}} $ denotes the element in the $ \mathbf{i} $-th row and $ \mathbf{j} $-th column of $ \mathbf{W} $. By applying the chain rule, we can express the computation of $ \mathbf{w_{ij}} $ as follows:
\begin{equation}
\label{loss_wij}
\mathbf{\frac{\partial loss}{\partial w_{ij}} =  \alpha \frac{\partial loss1}{\partial w_{ij}} +  (1-\alpha) \frac{\partial loss2}{\partial w_{ij}} + \beta \frac{\partial loss3}{\partial w_{ij}}}
\end{equation}

After obtaining the partial derivative of $ \mathbf{\frac{\partial loss}{\partial W}} $, we can update the optimal adjacency matrix $ \mathbf{W} $ using the following rule:
\begin{equation}
\label{w}
\mathbf{W = (1-\rho) W +  \rho \frac{\partial loss}{\partial W}}
\end{equation}
where $\rho$ is the learning rate hyperparameter we set during network training.

The detailed DGSD training algorithm is summarized in Algorithm~\ref{alg1}. For the ablation study, we perform it by removing the feature distillation or hierarchical distillation from the self-distillation method.

\renewcommand{\algorithmicrequire}{ \textbf{Input:}}     
\renewcommand{\algorithmicensure}{ \textbf{Output:}}    

\begin{algorithm}[H]
\caption{Training Algorithm for Optimal DGSD Model}
\begin{algorithmic}[1]
\REQUIRE Graph $ \mathbf{G=\{V, W\}} $ representing multi-channel EEG signals associated with multiple frequency bands, auditory spatial attention labels $\mathbf{y}$ corresponding to EEG, the number of layers of DGSD model $\mathbf{m}$, the number of Chebyshev polynomial order $\mathbf{K}$, the learning rate $\mathbf{\rho}$;
\ENSURE The optimal adjacency matrix $ \mathbf{W} $ and the optimal model parameters of DGSD;
\STATE Initialize the adjacency matrix $ \mathbf{W} $ and model parameters;
\STATE \textbf{repeat}
\STATE \hspace{0.3cm} Apply ReLU operation to normalize the elements $ \mathbf{w_{ij}} $ \\ \hspace*{9pt} in $ \mathbf{W} $, ensuring $ \mathbf{w_{ij} \geq 0} $ for all;
\STATE \hspace{0.3cm} Calculate and normalize the Laplacian matrix $ \mathbf{L} $; 
\STATE \hspace{0.3cm} Calculate the Chebyshev polynomials;
\STATE \hspace{0.3cm} Extract EEG signal features through each layer of \\ \hspace*{9pt} graph convolutional layers;
\STATE \hspace{0.3cm} Using average pooling to extract more representative \\ \hspace*{9pt} EEG features from each layer;
\STATE \hspace{0.3cm} Calculate binary classification probabilities using a \\ \hspace*{9pt} fully connected classifier for each layer;
\STATE \hspace{0.3cm} Calculate cross-entropy loss ($ \mathbf{loss1} $), feature distillation \\ \hspace*{9pt} loss ($ \mathbf{loss2} $), and hierarchical distillation loss ($ \mathbf{loss3} $) \\ \hspace*{9pt} using Eq.~(\ref{loss1}),~(\ref{loss2}), and~(\ref{loss3}) respectively;
\STATE \hspace{0.3cm} Calculate the loss function using Eq.~(\ref{loss});
\STATE \hspace{0.3cm} Update $ \mathbf{W} $ and other model parameters using back-\\ \hspace*{9pt} propagation;
\STATE \textbf{until} the iterations satisfy the predefined algorithm convergence condition;
\end{algorithmic}
\label{alg1}
\end{algorithm}

\section{Experiments}
\label{experiment}
In this section, we present the experimental details of DGSD. AAD datasets and EEG data preprocessing are briefly described in Section~\ref{data} and Section~\ref{prepro}, respectively. The evaluation metrics are described in Section~\ref{eva}, and implementation details and baseline descriptions are provided in Section~\ref{deta}. Additionally, EEG data and their attention direction labels are read from two original EEG public dataset files.

\subsection{AAD Datasets}
\label{data}
We validate our proposed method on the following two publicly available datasets, as shown in Table~\ref{dataset}, with detailed information about KUL and DTU available in \cite{ref51, ref52, ref53, ref54}.
\subsubsection{\textbf{KUL dataset}} 
This dataset contains 64-channel EEG data from 16 subjects, with an equal gender distribution (half male, half female). The data were recorded using the BioSemi ActiveTwo system with a sampling rate of 8192 Hz and an electrode layout conforming to the international 10/20 system. Auditory stimuli consisted of four Dutch short stories narrated by a male speaker. During the experiment, each subject was instructed to focus their attention on one of two competing male speakers narrating a story while ignoring the other. Auditory stimuli were presented at a volume of 60 dB through in-ear headphones and filtered with a low-pass cutoff frequency of 4 KHz. Each subject completed 20 trials, each lasting 6 minutes. There were two stimulus conditions: "HRTF" or "dry", resulting in a total of eight trials. Auditory stimuli were presented from the left at 90\textdegree\ and from the right at 90\textdegree\ by the two speakers. The presentation order was randomized across subjects. A total of 8 trials, each lasting 6 minutes, were collected for each subject, resulting in 48 minutes of EEG data. More detailed information about this dataset can be found in references\cite{ref51, ref52}.

\subsubsection{\textbf{DTU dataset}} 
This dataset contains 64-channel EEG data from 18 subjects. The data were recorded using the Biosemi system with a sampling rate of 512 Hz and an electrode layout conforming to the international 10/20 system. Auditory stimuli consisted of Danish audiobooks narrated by both male and female speakers. During the experiment, each subject was required to focus on one of two competing speakers (one male and one female) and ignore the other. To simulate low-reverberation conditions, the recordings of the two competing speakers were interfered with by six additional background speakers (three male and three female). The voices of the respective two speakers were presented from +60\textdegree\ and -60\textdegree\ relative to the subject as sound stimuli at a volume of 65 dB using ER-2 insert earphones with a sampling rate of 48 KHz. Each subject completed a total of 60 trials under three different conditions, with each trial lasting 50 seconds. Consequently, each subject collected 50 minutes of EEG data. For more detailed information about this dataset, please refer to references\cite{ref53, ref54}.

\begin{table*}[!t]
	\captionsetup{labelfont=bf}
	\caption{Details of the two datasets.}
	\label{dataset}
	\centering
	\renewcommand\arraystretch{1.2}
	\begin{tabular}{cccccc}
		\midrule
		\textbf{Dataset} & \begin{tabular}[c]{@{}c@{}}\textbf{Number of} \\ \textbf{subjects}\end{tabular} & \begin{tabular}[c]{@{}c@{}}\textbf{EEG duration}\\ \textbf{(per subject)}\end{tabular} & \textbf{Speakers}        & \textbf{Stimulus language}   & \begin{tabular}[c]{@{}c@{}}\textbf{Direction of} \\ \textbf{stimulus}\end{tabular} \\
		\midrule
		KUL     & 16                                                            & 48 min                                                               & Male            & Dutch short stories & $\pm$\ 90\textdegree                                                              \\
		DTU     & 18                                                            & 50 min                                                               & Male and female & Danish audiobooks   & $\pm$\ 60\textdegree                                                              		\\ \midrule
	\end{tabular}
\end{table*}

\subsection{EEG data preprocessing}
\label{prepro}
Preprocessing of EEG data is different from the EEG data processing in section~\ref{pro}. This section refers to a series of processing and correction of the raw EEG signals, which can improve the quality of the EEG signals and extract more effective features. Specifically, for the KUL dataset, the EEG signals are bandpass filtered from 0.1 Hz to 50 Hz, downsampled to 128 Hz, and then the brainwave data channels are normalized to ensure zero mean and unit variance across trials. For the DTU dataset, firstly, the line noise and harmonics at 50 Hz in the EEG signals are removed. Secondly, a resampling method based on the fast fourier transform (FFT) is used to downsample the EEG data to 128 Hz. Then, a joint decorrelation framework is used to remove eye artifacts, and a fourth-order forward Butterworth filter is applied to high-pass filter the EEG data at 1.0 Hz \cite{ref53}. Finally, each trial's EEG data are Z-normalized (also known as Z-score normalization) to ensure that they have unit variance and zero mean for each channel. The purpose of this process is to eliminate scale differences between different channels, allowing all EEG data from each channel to be compared and analyzed on the same scale.

\subsection{Evaluation metrics}
\label{eva}
Since our task involves detecting spatial direction, specifically left/right (i.e., 0/1), we can think of it as a binary classification task. In our study, we use two evaluation metrics to evaluate the model, the first is accuracy and standard deviation, and the second is precision and recall.

\subsubsection{\textbf{Accuracy and standard deviation}} 
Paired t-tests are used to compare the performance differences between two different models at a significance level of 0.05. The mean classification accuracy and standard deviation of all subjects in each dataset are computed under different time windows (0.5s, 1s, 2s, 5s).

\subsubsection{\textbf{Precision and recall}} 
In each dataset, under a chance level of 50\%, precision and recall are calculated for each subject, followed by the calculation of the mean precision and recall for all subjects.

\subsection{Implementation details}
\label{deta}
\subsubsection{\textbf{Training, validation and testing}} 
We implement the entire experiment using Python 3.7.0 and PyTorch 1.12.1. All experiments are conducted on NVIDIA GeForce RTX 3090 GPU. Our research is evaluated within the subject. After sliding window processing, the data of each subject is randomly divided into training, validation and test sets at a ratio of 8:1:1, and then each subject is trained and tested separately. The data in each table in the paper is the average of all subjects in the dataset. The random seed is set to 1111, batch size is set to 32, and the number of epochs is 200, with Adam used as the optimizer. To adapt to different datasets, the learning rates during the training process for KUL and DTU are set to 0.004 and 0.007, respectively. Two hyperparameters, $\alpha$ and $\beta$, are set to 0.7 and 0.3, respectively.

\subsubsection{\textbf{Baselines}} 
We use some baselines to evaluate the performance of the DGSD model. In order to ensure the fairness and validity of the performance comparison, the baseline models we compare are also tested on multiple datasets in their respective papers, which guarantees the generalization ability of the baseline models. All \cite{ref27, ref13} models are open-source implementations, and we indicate with an asterisk ($ ^{\ast} $) after the model name in the experimental results, where our replicated results are provided before "/", and the results reported in the baseline model papers are provided after "/". If there is only one result, it is the one we have reproduced because their paper does not contain any experiments on this dataset, and the baseline models we reproduce is performed under the same conditions as our model. As the implementation of \cite{ref8, ref55, ref5, ref30, ref56, ref57} is not yet formally open-sourced, their performance comes from their original papers, and we do not mark these results.

\section{Results}
\label{result}

\begin{table*}[t]
\captionsetup{labelfont=bf}
\caption{The proposed model achieves AAD accuracy (\%) on the KUL dataset compared to the baseline. In the experimental results of the baseline models marked with "*", the value before "/" represents the results we have reproduced, while the value after "/" represents the results reported in the baseline model paper. "-" notes that there are no experiments conducted on this dataset in the paper.}
\label{kul}
\centering
\renewcommand\arraystretch{1.2}
\begin{tabular}{ccccccc}
\midrule
\multirow{2}{*}{\textbf{Dataset}} &
  \multirow{2}{*}{\textbf{Model}} &
  \multirow{2}{*}{\textbf{Use auditory stimuli}} &
  \multicolumn{4}{c}{\textbf{Time Window}} \\ \cmidrule{4-7} 
                     &         &     & \textbf{0.5-second}    & \textbf{1-second}         & \textbf{2-second}         & \textbf{5-second}         \\ \midrule
\multirow{9}{*}{KUL} & S-R \cite{ref8}     & Yes & 53.9             & 58.1                & 61.3                & 67.5                \\
                     & CCA \cite{ref55}     & Yes & 55.4             & 59.2                & 62.4                & -                   \\
                     & DNN \cite{ref5}     & Yes & 64.9             & 70.7                & 74.5                & -                   \\
                     & BIAnet \cite{ref56}  & Yes & 84.1             & 84.4                & 88.1                & -                   \\
                     & CNN \cite{ref30}     & No  & 73.4             & 80.8                & 82.1                & 83.6                \\
                     & NI-AAD \cite{ref57}  & No  & 79.4             & 82.8                & 87.1                & 91.2                \\
                     & SSF-CNN$ ^{\ast} $ \cite{ref27} & No  & 80.5 ± 8.34 / - & 81.9 ± 9.86 / 81.7 & 87.3 ± 8.79 / 84.7 & 91.6 ± 7.40 / 90.5 \\
                     & MBSSFCC$ ^{\ast} $ \cite{ref13} & No  & 85.0 ± 7.50 / - & 88.8 ± 7.80 / 89.2 & 90.3 ± 7.62 / 91.5 & 92.8 ± 5.32 / 93.9 \\
 &
  \textbf{DGSD (ours)} &
  \textbf{No} &
  \textbf{86.3 ± 7.89} &
  \textbf{90.3 ± 7.29} &
  \textbf{93.3 ± 6.53} &
  \textbf{94.8 ± 4.61} \\ \hline
\end{tabular}
\end{table*}

\begin{table*}[t]
\captionsetup{labelfont=bf}
\caption{The proposed model achieves AAD accuracy (\%) on the DTU dataset compared to the baseline. In the experimental results of the baseline models marked with "*", the value before "/" represents the results we have reproduced, while the value after "/" represents the results reported in the baseline model paper. "-" notes that there are no experiments conducted on this dataset in the paper.}
\label{dtu}
\centering
\renewcommand\arraystretch{1.2}
\begin{tabular}{ccccccc}
\midrule
 &        &     & \multicolumn{4}{c}{\textbf{Time Window}} \\ \cmidrule{4-7} 
\multirow{-2}{*}{\textbf{Dataset}} &
  \multirow{-2}{*}{\textbf{Model}} &
  \multirow{-2}{*}{\textbf{Use auditory stimuli}} &
  \textbf{0.5-second} &
  \textbf{1-second} &
  \textbf{2-second} &
  \textbf{5-second} \\ \midrule
 & S-R \cite{ref8}    & Yes & -        & 51.8     & 55.3     & -       \\
 & CCA \cite{ref55}    & Yes & 51.2     & 53.5     & 58.9     & -       \\
 & DNN \cite{ref5}    & Yes & 56.8     & 61.7     & 62.8     & -       \\
 & BIAnet \cite{ref56} & Yes & 78.1     & 79.0     & 80.6     & -       \\
 & CNN \cite{ref30}    & No  & -        & 55.9     & 57.8     & 58.5    \\
 & NI-AAD \cite{ref57} & No  & 60.2     & 61.6     & 63.2     & 61.5    \\
 &
  SSF-CNN$ ^{\ast} $ \cite{ref27} &
  No &
  63.3 ± 6.42 / - &
  64.0 ± 7.21 / - &
  65.5 ± 7.47 / - &
  68.4 ± 13.89 / - \\
 &
  MBSSFCC$ ^{\ast} $ \cite{ref13} &
  No &
  71.3 ± 5.84 / - &
  75.2 ± 7.43 / 76.9 &
  78.7 ± 7.86 / 80.6 &
  80.2 ± 8.64 / 82.9 \\
\multirow{-9}{*}{DTU} &
  \textbf{DGSD (ours)} &
  \textbf{No} &
  \textbf{75.6 ± 6.72} &
  \textbf{79.6 ± 6.76} &
  \textbf{82.4 ± 6.86} &
  \textbf{85.6 ± 7.36} \\ \midrule
\end{tabular}
\end{table*}

\subsection{Low-latency DGSD} 

\begin{table*}[!t]
\captionsetup{labelfont=bf}
\caption{Metrics (accuracy and recall) for different time window lengths (0.5-second, 1-second, 2-second, 5-second) in the two datasets.  The value (\%) in this table represents the average for all subjects in each dataset.}
\label{metrics}
\centering
\renewcommand\arraystretch{1.2}
\begin{tabular}{cccccccccc}
\midrule
\multirow{3}{*}{\textbf{Dataset}} &
  \multirow{3}{*}{\textbf{Model}} &
  \multicolumn{8}{c}{\textbf{Time Window}} \\ \cmidrule{3-10} 
 &
   &
  \multicolumn{2}{c}{\textbf{0.5-second}} &
  \multicolumn{2}{c}{\textbf{1-second}} &
  \multicolumn{2}{c}{\textbf{2-second}} &
  \multicolumn{2}{c}{\textbf{5-second}} \\
 &  & \textbf{precision} & \textbf{recall} & \textbf{precision} & \textbf{recall} & \textbf{precision} & \textbf{recall} & \textbf{precision} & \textbf{recall} \\ \midrule
\multirow{3}{*}{KUL} &
  SSF-CNN$ ^{\ast} $ \cite{ref27} &
  81.5 &
  78.7 &
  82.1 &
  81.8 &
  86.6 &
  89.3 &
  92.5 &
  90.8 \\
 &
  MBSSFCC$ ^{\ast} $ \cite{ref13} &
  85.2 &
  84.9 &
  89.1 &
  88.6 &
  90.1 &
  90.9 &
  94.1 &
  91.6 \\
 &
  \textbf{DGSD (ours)} &
  \textbf{86.8} &
  \textbf{85.3} &
  \textbf{89.4} &
  \textbf{89.3} &
  \textbf{93.4} &
  \textbf{93.2} &
  \textbf{94.6} &
  \textbf{95.4} \\ \midrule
\multirow{3}{*}{DTU} &
  SSF-CNN$ ^{\ast} $ \cite{ref27} &
  63.5 &
  60.2 &
  64.7 &
  58.1 &
  66.6 &
  61.4 &
  69.6 &
  64.9 \\
 &
  MBSSFCC$ ^{\ast} $ \cite{ref13} &
  70.9 &
  70.6 &
  73.5 &
  73.2 &
  76.0 &
  80.4 &
  79.1 &
  76.9 \\
 &
  \textbf{DGSD (ours)} &
  \textbf{71.3} &
  \textbf{77.0} &
  \textbf{79.6} &
  \textbf{78.6} &
  \textbf{81.2} &
  \textbf{81.7} &
  \textbf{83.6} &
  \textbf{84.1} \\ \midrule
\end{tabular}
\end{table*}

\begin{figure*}[!t]
\centering
\subfloat[\footnotesize KUL dataset]{\includegraphics[width=15.5cm,height=4.7cm]{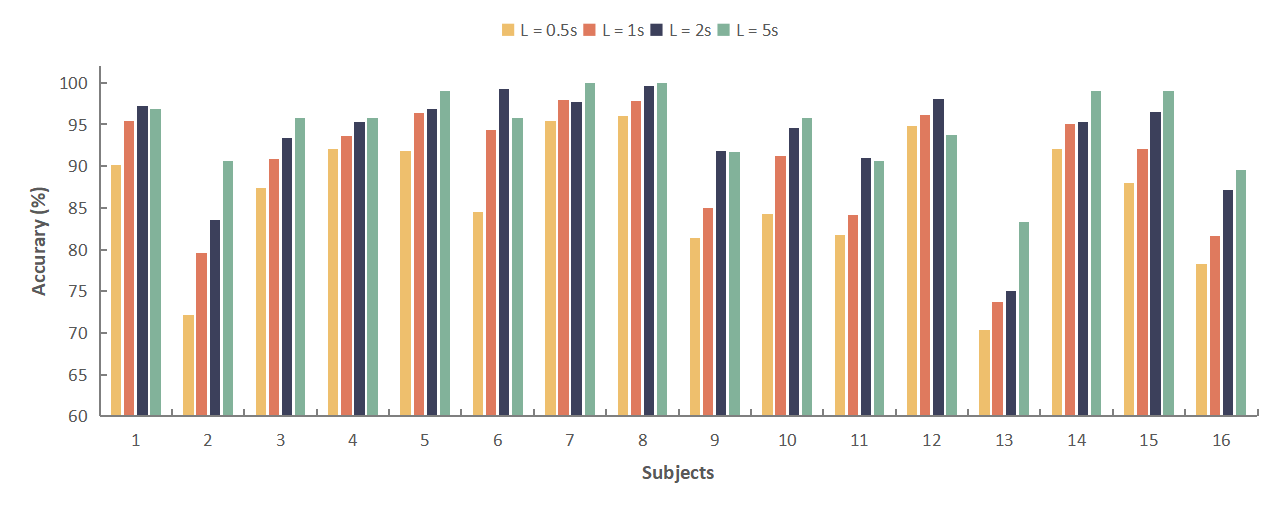}%
\label{sub1}}
\hfil
\subfloat[\footnotesize DTU dataset]{\includegraphics[width=15.5cm,height=4.7cm]{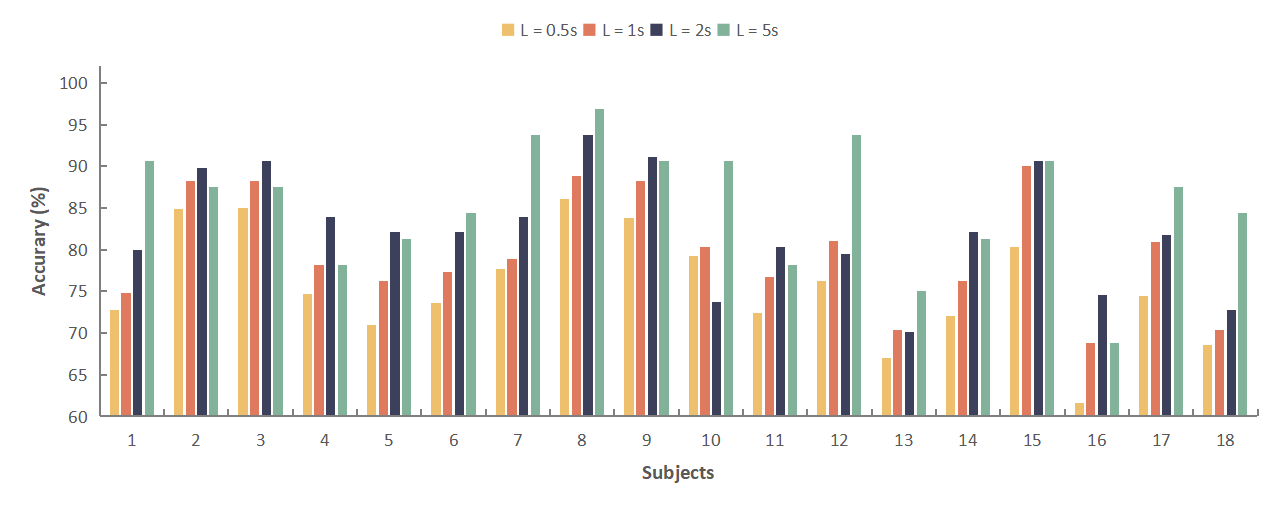}%
\label{sub2}}
\caption{Detection accuracy (\%) of the DGSD model implemented on each subject in the KUL and DTU datasets for different decision time window lengths (0.5-second, 1-second, 2-second, 5-second). Sort the horizontal axis by the subject IDs. (a) KUL dataset. (b) DTU dataset.}
\label{sub}
\end{figure*}

To evaluate the feasibility of the DGSD model in practical applications, research is being conducted on the proposed DGSD model with four different time windows. The detection accuracy of the DGSD model on two datasets is reported in Table~\ref{kul} and Table~\ref{dtu}, covering time windows ranging from relatively short durations of 0.5-second to relatively longer durations of 5-second. Table~\ref{metrics} also presents the metrics (precision and recall) of the DGSD model for four time windows on both datasets. Additionally, Fig.~\ref{sub} illustrates the performance of the DGSD model with different time windows for each subject in the two datasets. The horizontal axis is sorted by subject ID, and the vertical axis represents the accuracy of auditory spatial attention detection (starting at 60\%). It is observed that in the KUL dataset (Fig.~\ref{sub1}) and the DTU dataset (Fig.~\ref{sub2}).

From Table~\ref{kul}, it can be observed that in the KUL dataset, the DGSD exhibits excellent auditory attention detection performance under 1-second time window (mean: 90.3\%, SD: 7.29\%), 2-second time window (mean: 93.3\%, SD: 6.53\%), and 5-second time window (mean: 94.8\%, SD: 4.61\%). Moreover, as the time window shortens, the performance of the DGSD model under the 0.5-second time window (mean: 86.3\%, SD: 7.89\%) decreases with the reduction of EEG signal information, but still maintains a very high detection performance. It can be inferred from the research results that as the time window increases, the detection accuracy of the DGSD model significantly improves, which is consistent with the findings of \cite{ref13, ref56, ref57}.

From Table~\ref{dtu}, it can be observed that in the DTU dataset, the DGSD model achieves average accuracies of 75.6\% (SD: 6.72\%), 79.6\% (SD: 6.76\%), 82.4\% (SD: 6.86\%), and 85.6\% (SD: 7.36\%) for time windows of 0.5-second, 1-second, 2-second, and 5-second, respectively. Similar to the results on the KUL dataset, the trend in results is consistent, indicating an improvement in auditory spatial attention detection accuracy with the increase in time window size.

Table~\ref{metrics} also showcases the metrics (precision and recall) of the DGSD model with four time windows on both datasets. As the time window increases, these two metrics also improve. It can be observed that in the KUL dataset (Fig.~\ref{sub1}) and the DTU dataset (Fig.~\ref{sub2}), as the time window L increases from 0.5-second to 5-second, although there are some exceptions, the detection accuracy of most subjects gradually rises. This suggests that as L becomes longer, more information is captured in the EEG signals after sliding window processing, allowing our DGSD model to extract more useful features for auditory attention detection.

However, a noteworthy observation is that the detection performance of the DGSD model on the DTU dataset is lower compared to the KUL dataset, aligning with the findings in studies\cite{ref13, ref30, ref56, ref57}. Through a analysis of the publicly available descriptions of these two datasets, we suggest that this may be due to the direction of the auditory stimulus or the gender of the speaker. The primary distinctions between these datasets are as follows:

\subsubsection{Directional bias of attention (±90° vs ±60°)}In the DTU dataset, the two auditory stimuli are distributed at ±60° angles relative to the subjects, while in the KUL dataset, they are distributed at ±90° angles. Subjects might naturally exhibit a more pronounced attention bias towards the ±90° direction, making the auditory stimuli from the KUL dataset potentially more attention-grabbing and thus yielding higher performance.
\subsubsection{Gender-related influence (Male \& female vs Male)}The auditory stimuli in the DTU dataset are presented by both male and female speakers, whereas in the KUL dataset, they are presented only by male speakers. Variations in tone and frequency may exist between male and female voices. The auditory stimuli from both male and female speakers in the DTU dataset could possess distinct voice characteristics. This divergence might impact subjects' attention biases towards stimuli of different genders.

Therefore, we consider that researching the DTU dataset could present more challenges.

\subsection{Ablation study} 

\begin{table*}[!t]
\captionsetup{labelfont=bf}
\caption{AAD accuracy and standard deviation (\%) obtained
from the ablation study of loss functions on DGCN.}
\label{abl}
\centering
\renewcommand\arraystretch{1.2}
\begin{tabular}{cccccc}
\midrule
\multirow{2}{*}{\textbf{Dataset}} & \multirow{2}{*}{\textbf{Loss Function}} & \multicolumn{4}{c}{\textbf{Time Window}}                                                  \\ \cmidrule{3-6} 
                                  &                                         & \textbf{0.5-second}        & \textbf{1-second}          & \textbf{2-second}          & \textbf{5-second}          \\ \midrule
\multirow{4}{*}{KUL}              & loss1 (DGCN)                                & 85.1 ± 8.18          & 89.0 ± 7.22          & 92.6 ± 6.49          & 93.8 ± 5.49          \\
                                  & loss1 + loss2                                 & 85.8 ± 7.49          & 89.5 ± 7.60          & 92.2 ± 7.06          & 93.8 ± 5.82          \\
                                  & loss1 + loss3                                 & 86.0 ± 8.14          & 89.5 ± 8.23          & 92.1 ± 7.49          & 94.6 ± 5.08          \\
                                  & \textbf{loss1 + loss2 + loss3 (DGSD)}             & \textbf{86.3 ± 7.89} & \textbf{90.3 ± 7.29} & \textbf{93.3 ± 6.53} & \textbf{94.8 ± 4.61} \\ \midrule
\multirow{4}{*}{DTU}              & loss1 (DGCN)                                & 73.8 ± 6.95          & 79.1 ± 7.12          & 81.3 ± 6.68          & 83.6 ± 9.26          \\
                                  & loss1 + loss2                                 & 73.1 ± 8.55          & 78.9 ± 6.13          & 81.7 ± 7.48          & 84.9 ± 6.36          \\
                                  & loss1 + loss3                                 & 73.0 ± 9.03          & 78.8 ± 6.36          & 80.9 ± 6.90          & 83.7 ± 7.55          \\
                                  & \textbf{loss1 + loss2 + loss3 (DGSD)}             & \textbf{75.6 ± 6.72} & \textbf{79.6 ± 6.76} & \textbf{82.4 ± 6.86} & \textbf{85.6 ± 7.36} \\ \midrule
\end{tabular}
\end{table*}

In order to evaluate the effectiveness of self-distillation (SD) in our DGSD model, we conduct an ablation study on the loss functions, investigating the impact of different combinations of loss functions. During the study, considering the components of the loss function "loss" in our approach: loss1 (cross-entropy loss), loss2 (feature distillation loss), and loss3 (hierarchical distillation loss), we employ loss1 as the primary loss and examine the performance when combined with loss2 or loss3 at specific proportions. Specific results are shown in Table~\ref{abl}. It can be observed that:

\subsubsection{\textbf{Only DGCN}}When using only DGCN (i.e., using loss1 as the sole loss function), its detection accuracy across different time windows surpasses the baseline models in Table~\ref{kul} and Table~\ref{dtu}. This suggests that the DGCN can suit to use graph structure to represent with a nature of non-Euclidean EEG signals, and can effectively extract and utilize the feature information in EEG signals, resulting in higher detection accuracy.

\subsubsection{\textbf{Combination of loss functions}}From experimental results, it can be seen that combining loss1 with either of the other two loss functions has little impact on detection accuracy. However, when all three loss functions are combined proportionally, our DGSD model outperforms DGCN by approximately 1\%. This indicates that the combination of the three loss functions is the optimal choice for the task, as they collaborate to provide enhanced performance and effectiveness. The experimental results show that our SD approach, which combines feature distillation and hierarchical distillation, pays more attention to the multi-level representation of features and labels, and can use the features and classification results of the deepest network to guide the learning of shallow networks, so that shallow networks are more helpful to extract the features of auditory spatial attention and get the correct classification results. It is helpful to improve the classification accuracy of auditory attention detection.

\subsection{Selection of hyperparameters} 

\begin{figure*}[!t]
\centering
\subfloat[\footnotesize $\alpha$ = 0.7]{\includegraphics[width=7cm,height=5cm]{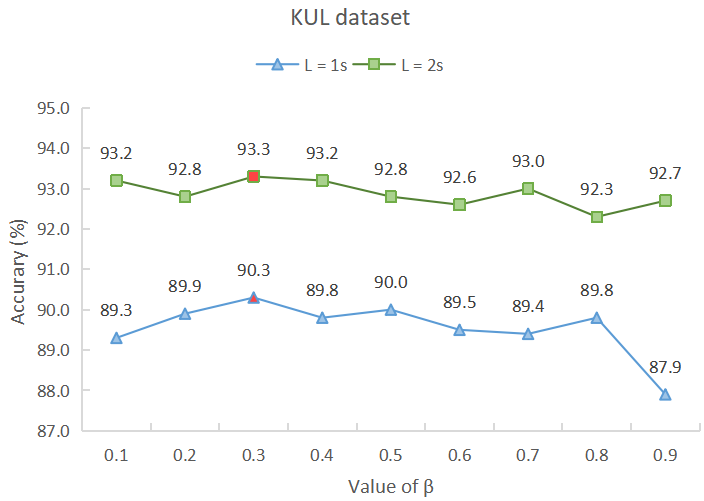}%
\label{zhe1}}
\hfil 
\subfloat[\footnotesize $\beta$ = 0.3]{\includegraphics[width=7cm,height=5cm]{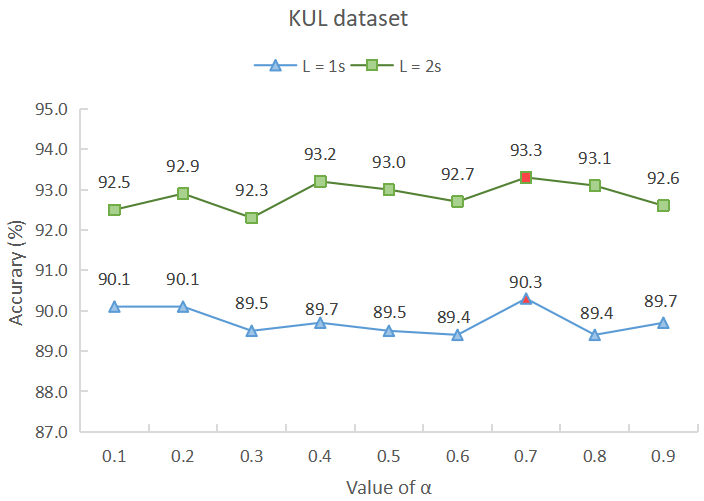}%
\label{zhe2}}
\hfil
\subfloat[\footnotesize $\alpha$ = 0.7]{\includegraphics[width=7cm,height=5cm]{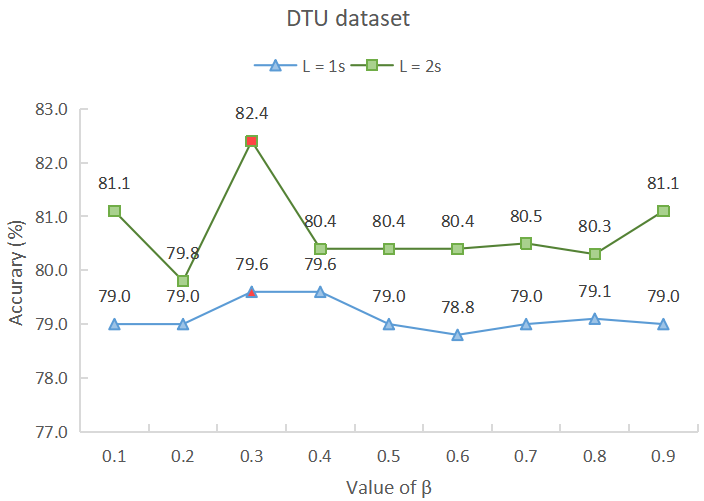}%
\label{zhe3}}
\hfil 
\subfloat[\footnotesize $\beta$ = 0.3]{\includegraphics[width=7cm,height=5cm]{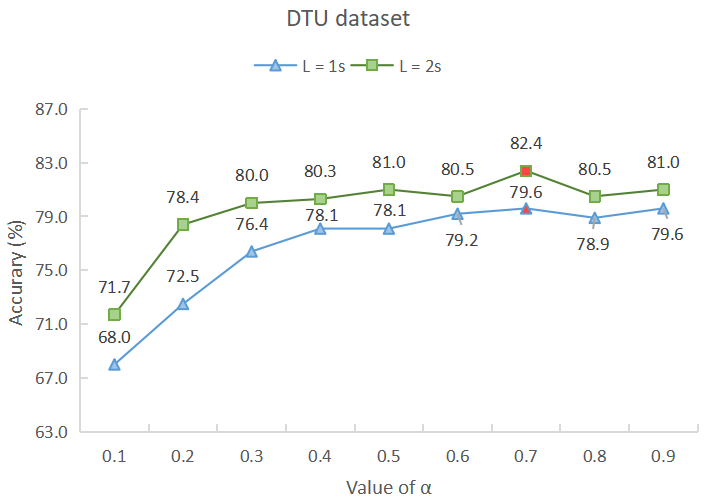}%
\label{zhe4}}
\caption{In two public datasets (KUL, DTU), the impact of different parameter combinations on accuracy (\%). Fig.~\ref{zhe1} and Fig.~\ref{zhe3} depict the accuracy under varying values of $\beta$ when $\alpha$ is set to 0.7. Fig.~\ref{zhe2} and Fig.~\ref{zhe4} illustrate the accuracy under different values of $\alpha$ when $\beta$ is fixed at 0.3. (a) $\alpha$ = 0.7. (b) $\beta$ = 0.3 (c) $\alpha$ = 0.7. (d) $\beta$ = 0.3.}
\label{zhe}
\end{figure*}

Studies show that it takes approximately 1-second to 2-second for a normal person to shift attention to another speaker\cite{ref47}. Therefore, suitable parameter combinations for Equation~\ref{loss} are being sought through parameter tuning of hyperparameters $\alpha$ and $\beta$ within 1-second to 2-second time windows to achieve optimal detection accuracy. The accuracy (\%) under 1-second and 2-second time windows for different parameter combinations is depicted in Fig.~\ref{zhe}, and extensive experiments are conducted by separately fixing the values of $\alpha$ and $\beta$ to obtain these results. Overall, these experiments can be classified into two types:

\subsubsection{\textbf{$\alpha$ = 0.7}}On the KUL dataset (Fig.~\ref{zhe1}) and the DTU dataset (Fig.~\ref{zhe3}), we fix $\alpha$ at 0.7 and vary the value of $\beta$.
\subsubsection{\textbf{$\beta$ = 0.3}}On the KUL dataset (Fig.~\ref{zhe2}) and the DTU dataset (Fig.~\ref{zhe4}), we fix $\beta$ at 0.3 and vary the value of $\alpha$.

It can be seen that, regardless of the subfigure, the detection accuracy of the two datasets in the 1-second and 2-second time windows reaches the optimum when $\alpha$ is 0.7 and $\beta$ is 0.3. This indicates that our model can effectively utilize spatial information (both local and global) in EEG signals for auditory attention detection under the above-mentioned hyperparameter settings.

\section{Discussion}
\label{discussion}
We believe that our proposed DGSD model not only effectively represents the channels of EEG signals but also adeptly extracts and classifies the relevant auditory attention information within the EEG signals. In this section, we begin by comparing the performance of our proposed DGSD model with models incorporating auditory stimuli. Subsequently, we also compare its performance with models without auditory stimuli, as depicted in Table~\ref{kul} and Table~\ref{dtu}. Moreover, regarding the reproduction of state-of-the-art open-source models, namely SSF-CNN\cite{ref27} and MBSSFCC\cite{ref13}, we analyze the precision and recall of these models in comparison with the DGSD model under various time windows for the KUL and DTU datasets, as presented in Table~\ref{metrics}. We further analyze the trainable parameter counts of the aforementioned models as well as our DGSD model, which are detailed in Table~\ref{para}. Lastly, we provide an interpretation of the results from our conducted ablation experiments on the loss functions. These experimental outcomes are available in Table~\ref{abl}.

\subsection{Performance comparison} 
\subsubsection{\textbf{DGSD vs Models (use auditory stimuli)}}
We compare the DGSD model with models that utilize auditory stimuli, which are models with "Use auditory stimuli" values set to "Yes" in Table~\ref{kul} and Table~\ref{dtu} (S-R\cite{ref8}, CCA\cite{ref55}, DNN\cite{ref5}, BIAnet\cite{ref56}). The results for these models are derived from their respective papers, where a "-" indicates that the experiment for that specific time window is not conducted in the model paper. While this comparison is conducted under different AAD paradigms, their objectives are the same — to identify and enhance auditory stimuli that the listener pays attention to, while attenuating other auditory stimuli that the listener neglects.

From Table~\ref{kul}, we observe that on the KUL dataset, DGSD accuracy (mean\_0.5-second: 86.3\%, mean\_1-second: 90.3\%, mean\_2-second: 93.3\%) for time windows of 0.5-second, 1-second, and 2-second significantly surpasses other models using auditory stimuli. Compared to the state-of-the-art auditory stimulus model, BIAnet (mean\_0.5-second: 84.1\%, mean\_1-second: 84.4\%, mean\_2-second: 88.1\%), the DGSD model achieves an average accuracy improvement of 2.2\%, 5.9\%, and 5.2\% respectively. From Table~\ref{dtu}, the effectiveness of the model is also verified on the DTU dataset. Although the DGSD model's accuracy is lower than BIAnet for the 0.5-second time window, it outperforms the BIAnet model for the 1-second and 2-second time windows. 

The experimental results demonstrate that our proposed DGSD model can achieve higher AAD accuracy without utilizing auditory stimuli, making it more suitable for real-life scenarios.

\subsubsection{\textbf{DGSD vs Models (do not use auditory stimuli)}}
We compare models that share the same AAD paradigm with the DGSD model. These models do not require auditory stimuli as inputs, which is more aligned with practical applications. The models not using auditory stimuli are those listed with "Use auditory stimuli" values set to "No" in Table~\ref{kul} and Table~\ref{dtu} (CNN\cite{ref30}, NI-AAD\cite{ref57}, SSF-CNN\cite{ref27}, MBSSFCC\cite{ref13}), where the results for CNN and NI-AAD models are derived from their respective papers. We focus on comparing the DGSD model with the state-of-the-art open-source SSF-CNN and MBSSFCC models that we have reproduced. The comparison is structured as our experiment results followed by the results from the respective papers.

The results of this comparison on the KUL dataset can be seen in Table~\ref{kul} (p \textless 0.001). In the 1-second time window, the DGSD model achieves significantly higher detection accuracy (mean: 90.3\%, SD: 7.29\%) compared to SSF-CNN and MBSSFCC models, with average improvements of 8.4\% and 1.5\% respectively. In the 2-second time window, the DGSD model's detection accuracy (mean: 93.3\%, SD: 6.53\%) is on average 6.0\% and 3.0\% higher than SSF-CNN and MBSSFCC. Similarly, the validation on the DTU dataset, as shown in Table~\ref{dtu} (p \textless 0.001), demonstrates that in the 1-second time window, the DGSD model achieves significantly higher detection accuracy (mean: 79.6\%, SD: 6.76\%) compared to SSF-CNN and MBSSFCC models, with average improvements of 15.6\% and 4.4\% respectively. In the 2-second time window, the DGSD model's detection accuracy (mean: 82.4\%, SD: 6.86\%) is on average 16.9\% and 3.7\% higher than SSF-CNN and MBSSFCC.

It is evident that even without the use of auditory stimuli, our DGSD model achieves optimal classification detection results. This outcome underscores the effectiveness of our model in auditory spatial attention detection. Additionally, we compute the precision and recall of DGSD, SSF-CNN, and MBSSFCC under different time windows, as shown in Table~\ref{metrics}. Across different time windows, the DGSD model outperforms SSF-CNN and MBSSFCC models on both metrics. This shows that the DGSD model is more accurate than the other two models in predicting the left-right spatial direction.

Finally, we compare the trainable parameter counts of the proposed DGSD model, SSF-CNN, and MBSSFCC. As shown in Table~\ref{para}, the DGSD model achieves higher classification accuracy compared to SSF-CNN and MBSSFCC models, while requiring approximately 28 times fewer parameters than SSF-CNN and 100 times fewer parameters than MBSSFCC. This indicates that under the same setting, our model offers faster training speed and reduced storage requirements. This suggests that our DGSD method is more suitable for practical applications such as hearing AIDS, as it is faster at the same time with high accuracy.

\begin{table}[!t]
	\captionsetup{labelfont=bf}
	\centering
	\caption{The trainable parameter counts of our proposed DGSD model and two superior baseline models are being presented. "M" represents the "million" scale in numerical magnitude, which is equivalent to $ 10^{6} $.}
	\label{para}
		\centering
		\renewcommand\arraystretch{1.2}
				\begin{tabular}{cc}
					\midrule
					\textbf{Model}     & \textbf{Trainable parameters} \\ \midrule
					SSF-CNN$ ^{\ast} $ \cite{ref27}   & 4.21M                \\
					MBSSFCC$ ^{\ast} $ \cite{ref13}  & 16.87M               \\
					\textbf{DGSD (ours)} & \textbf{0.15M}                \\ \midrule
			\end{tabular}
\end{table}

\subsection{Analysis of Self-distillation} 

Our self-distillation consists of feature distillation and hierarchical distillation. As indicated by Table~\ref{abl}, when using only feature distillation (loss2) or hierarchical distillation (loss3) individually, the model performance is moderate. However, when they are combined, the effect improves. We believe this outcome is due to the following reasons:

\subsubsection{\textbf{loss1 is combined with loss2/loss3}}When using feature distillation alone, since there is no label assistance, the extracted features related to auditory attention may not necessarily be what we desire. When using hierarchical distillation alone, as it lacks the support of features, the obtained labels may not be desirable.

\subsubsection{\textbf{loss1 is combined with loss2 and loss3}}The combination of both compensates for their respective shortcomings. Feature distillation, through multi-level feature propagation, corrects inaccurate labels that hierarchical distillation might produce. The labels obtained from hierarchical distillation effectively guide feature distillation, emphasizing the extraction of features related to auditory attention direction.

\section{Conclusion}
\label{conclusion}
This paper introduces a DGSD model that combines self-distillation with dynamic graph convolutional networks. This model does not require auditory stimuli as input and relies solely on EEG signals for auditory spatial attention detection, making it more practical in real-world scenarios. Furthermore, the DGSD model effectively extracts crucial feature information related to auditory spatial attention from the EEG signals, with the self-distillation strategy enhancing the detection performance. The experimental results show that the DGSD model not only outperforms the linear model, but also outperforms the advanced reproducible nonlinear model while reducing the number of trainable parameters by about 100 times, demonstrating the effectiveness of our model in detecting auditory spatial attention. In summary, our DGSD model enhances the performance of EEG-based auditory attention detection and opens up endless possibilities for the development of various hearing devices in the future. Our research is conducted based on within-subject, and there is a lack of cross-subject research. Future work will be extended to cross-subject studies, which will help verify the consistency and robustness of the model.


\begin{IEEEbiography}[{\includegraphics[width=1.1in,height=1.25in,clip,keepaspectratio]{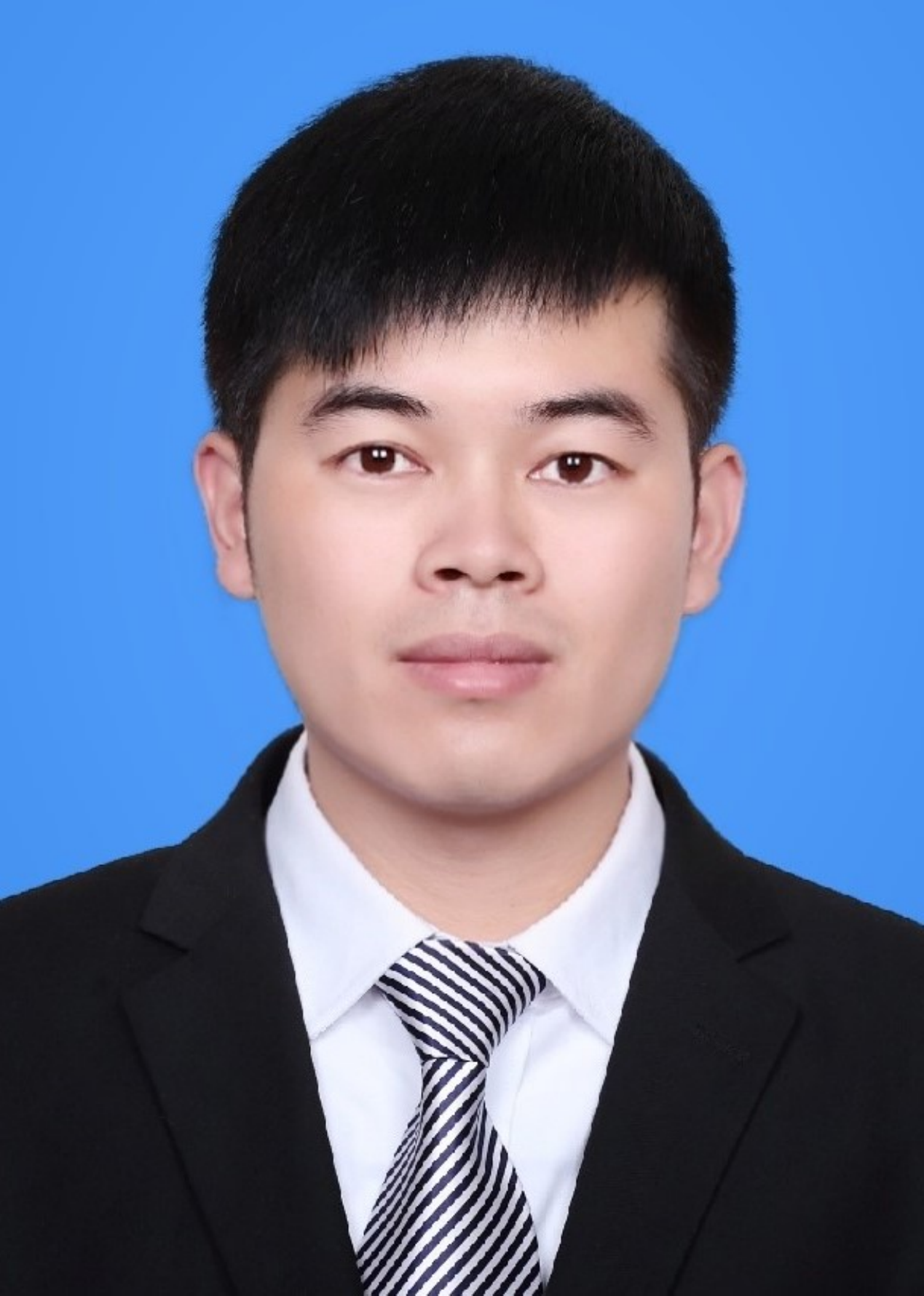}}]{Cunhang Fan}
	received the Ph.D degree with the National Laboratory of Pattern Recognition (NLPR), Institute of Automation, Chinese Academy of Sciences (CASIA), Beijing, China, in 2021, and the B.S. degree from the Beijing University of Chemical Technology (BUCT), Beijing, China, in 2016. He is currently a associate professor with the School of Computer Science and Technology, Anhui University, Heifei, China. His current research interests include auditory attention detection, speech enhancement, speech recognition and speech processing.
\end{IEEEbiography}

\begin{IEEEbiography}[{\includegraphics[width=1.1in,height=1.25in,clip,keepaspectratio]{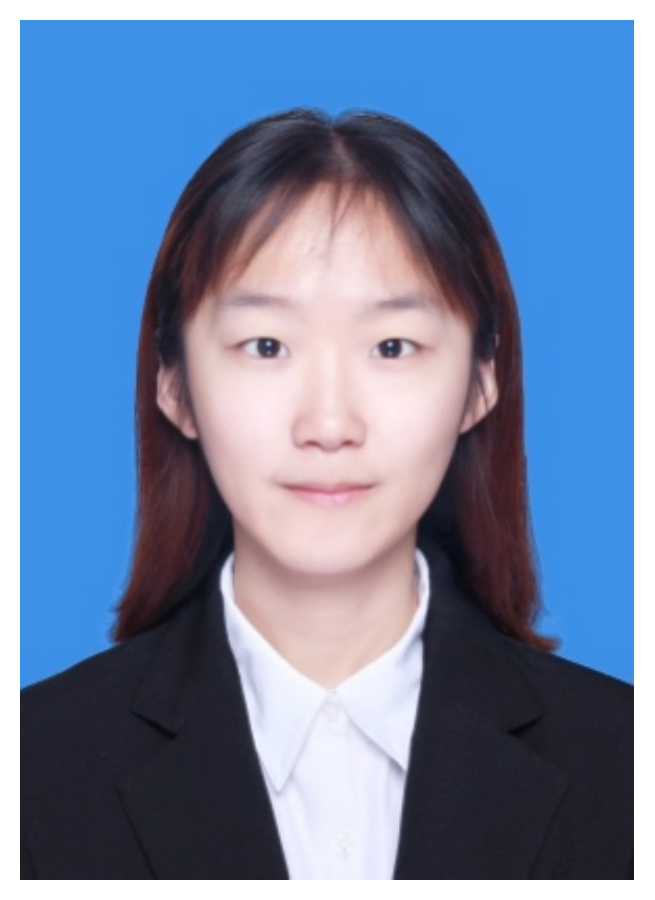}}]{Hongyu Zhang}
	graduated from Jinan University in 2022 with a Bachelor's degree in Computer Science and Technology. She is currently pursuing a Master's degree at the School of Computer Science and Technology, Anhui University. Her current research interests include auditory attention detection and brain-computer interfaces.
\end{IEEEbiography}

\begin{IEEEbiography}[{\includegraphics[width=1.1in,height=1.25in,clip,keepaspectratio]{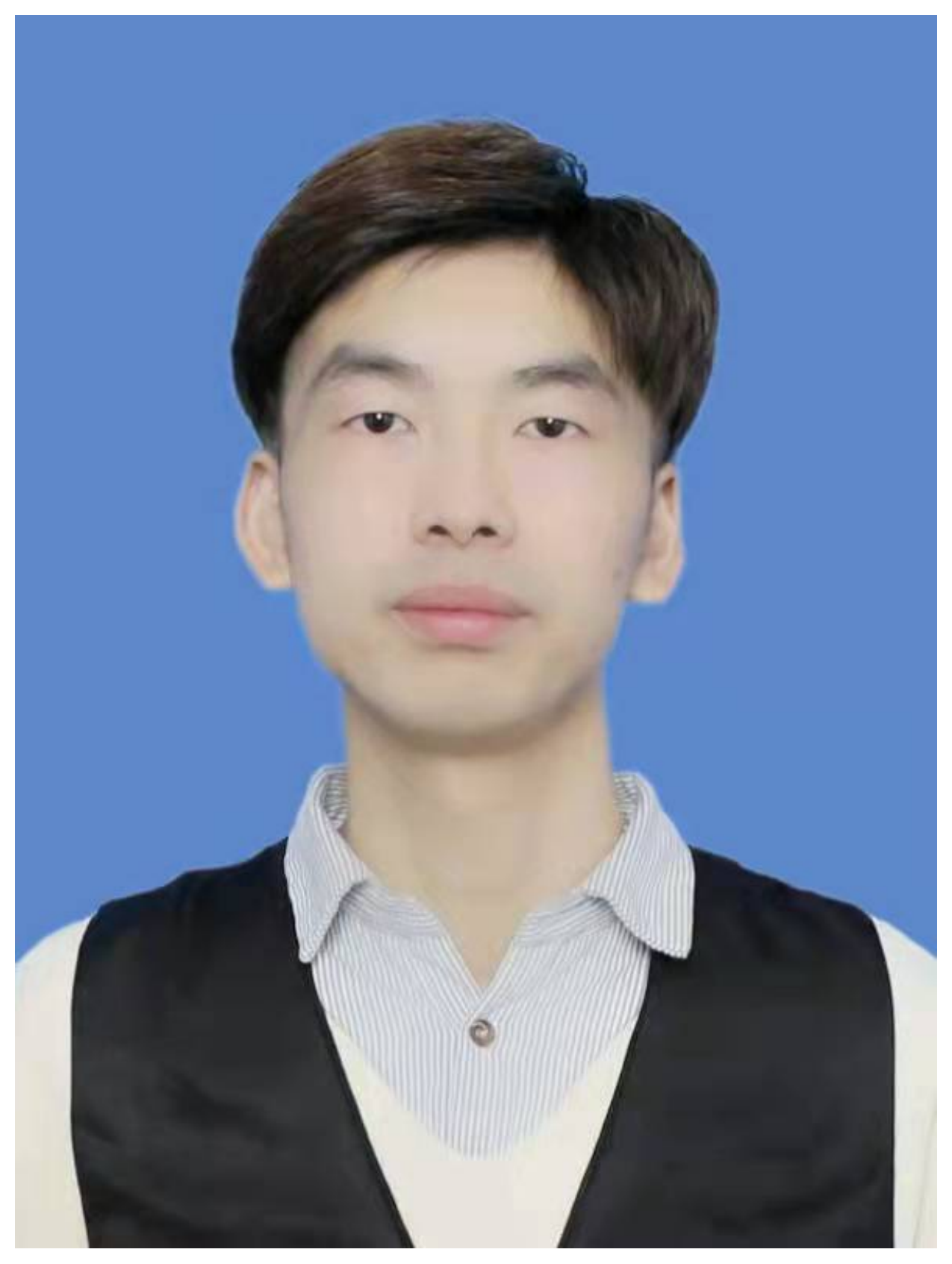}}]{Jun Xue}
	received the B.S. degree from the Anhui Science And Technology University, in 2020, and the M.S. degree at Anhui University from 2021 to the present. His research interests include: Fake speech detection, Knowledge distillation and Self-supervised learning.
\end{IEEEbiography}

\begin{IEEEbiography}[{\includegraphics[width=1.1in,height=1.25in,clip,keepaspectratio]{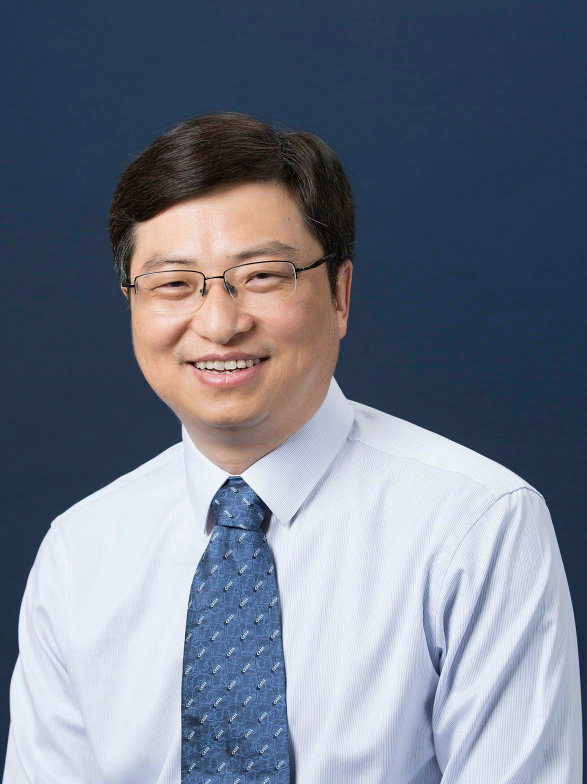}}]{Jianhua Tao}
	(Senior Member, IEEE) received the M.S. degree from Nanjing University, Nanjing, China, in 1996, and the Ph.D. degree from Tsinghua University, Beijing, China, in 2001. He is currently a Professor with Department of Automation, Tsinghua University, Beijing, China. He has authored or coauthored more than 300 papers on major journals and proceedings including the IEEE TASLP, IEEE TAFFC, IEEE TIP, IEEE TSMCB, Information Fusion, etc. His current research interests include speech recognition and synthesis, affective computing, and pattern recognition. He is the Board Member of ISCA, the chairperson of ISCA SIG-CSLP, the Chair or Program Committee Member for several major conferences, including Interspeech, ICPR, ACII, ICMI, ISCSLP, etc. He was the subject editor for the Speech Communication, and is an Associate Editor for Journal on Multimodal User Interface and International Journal on Synthetic Emotions. He was the recipient of several awards from the important conferences, including Interspeech, NCMMSC, etc.
\end{IEEEbiography}

\begin{IEEEbiography}[{\includegraphics[width=1.1in,height=1.25in,clip,keepaspectratio]{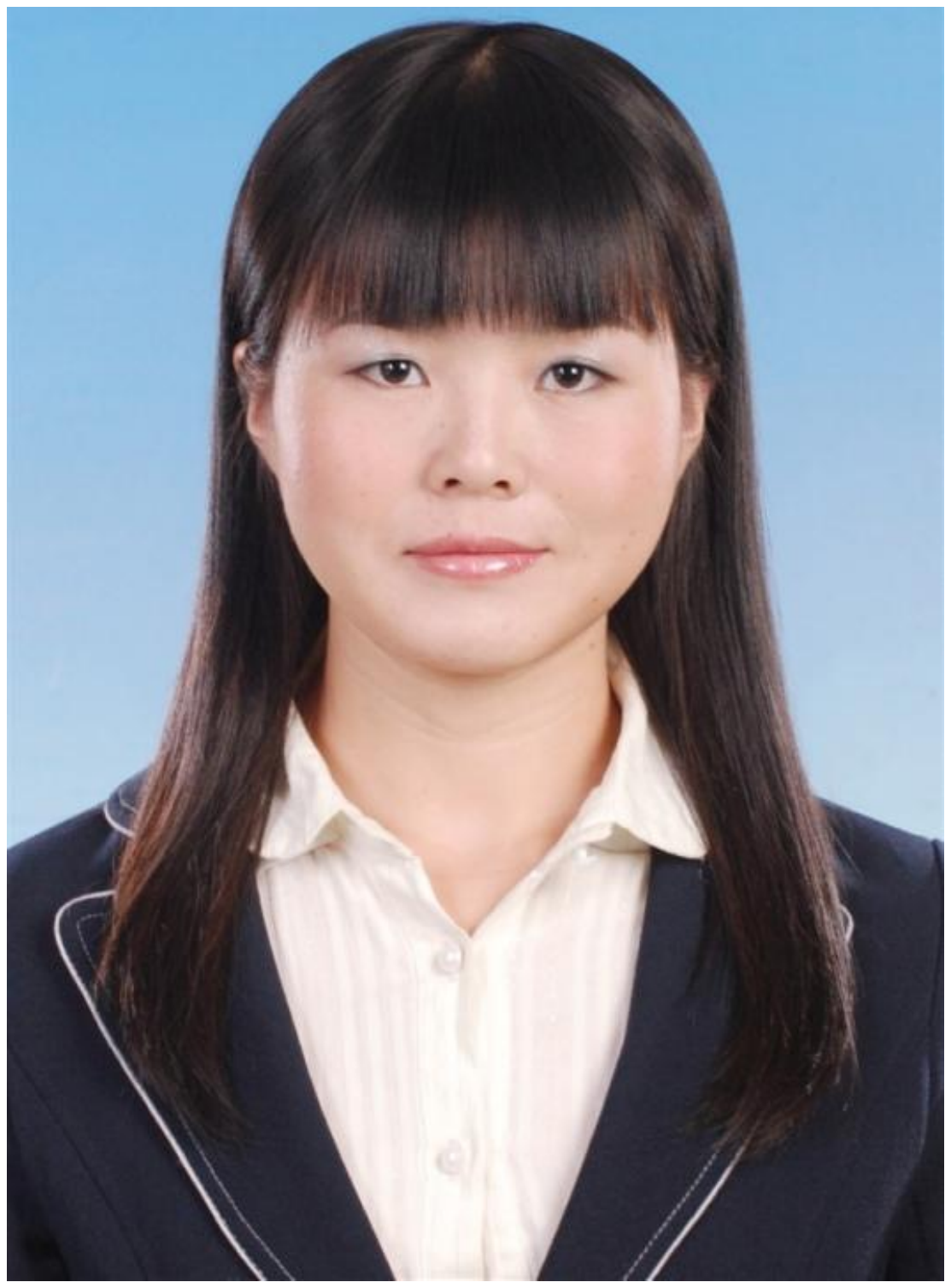}}]{Jiangyan Yi}
	received the Ph.D. degree from the University of Chinese Academy of Sciences, Beijing, China, in 2018, and the M.A. degree fromthe Graduate School of Chinese Academy of Social Sciences, Beijing, China, in 2010. She was a Senior R\&D Engineer with Alibaba Group during 2011 to 2014. She is currently an Assistant Professor with the National Laboratory of Pattern Recognition, Institute of Automation, Chinese Academy of Sciences, Beijing, China. Her current research interests include speech processing, speech recognition, distributed computing, deep learning, and transfer learning. 
\end{IEEEbiography}

\begin{IEEEbiography}[{\includegraphics[width=1.1in,height=1.25in,clip,keepaspectratio]{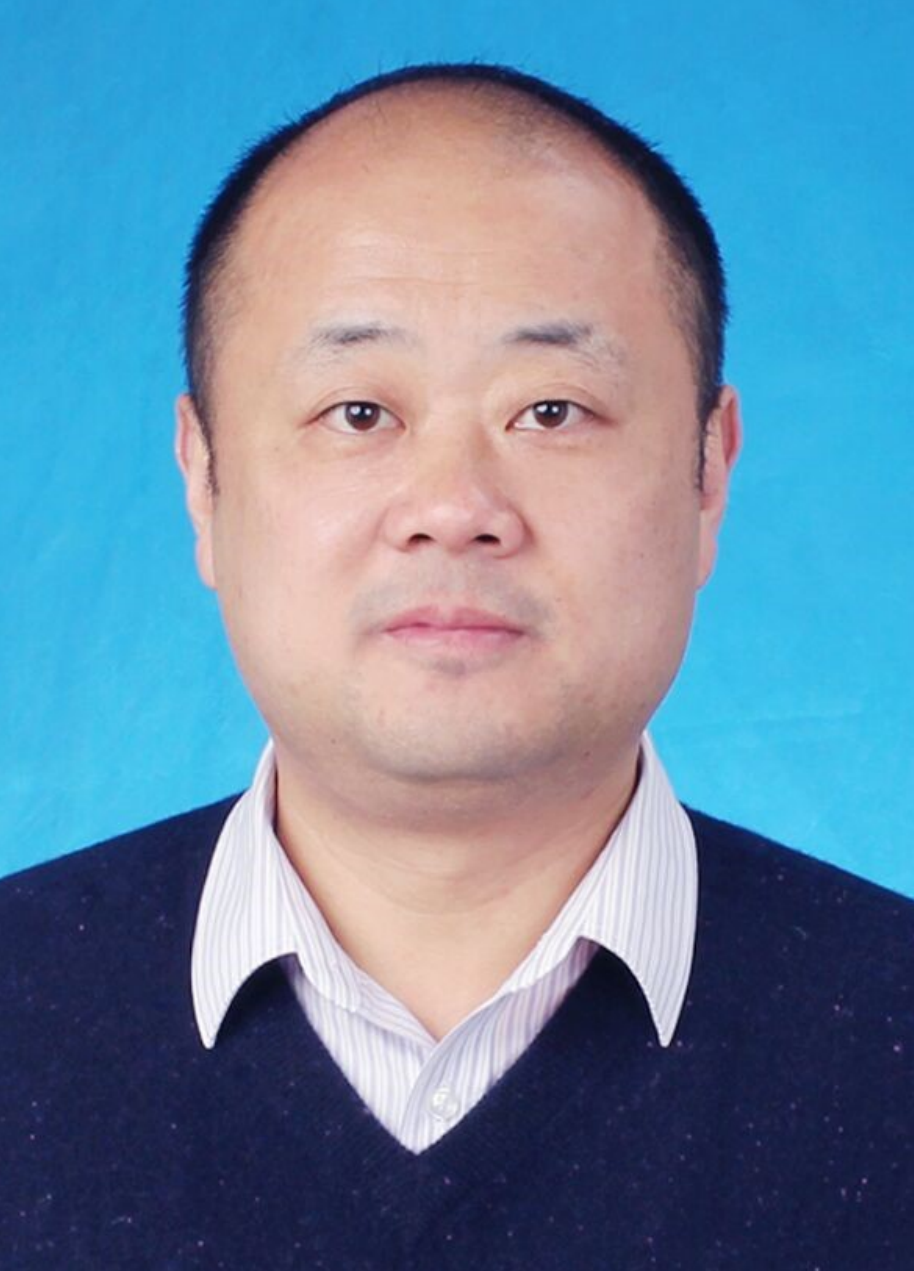}}]{Zhao Lv}
	received his Ph.D. degree in Computer Application Technology from Anhui University, Hefei, China, in 2011. He was a visiting scholar with the University of Utah, Salt Lake City, USA, from 2017 to 2018. He is currently a professor in the School of Computer Science and Technology at Anhui University, Hefei, China. His research interests include intelligent information processing and pattern recognition regarding biomedical signal (EEG, EOG, etc.) as well as speech signal processing.  
\end{IEEEbiography}

\begin{IEEEbiography}[{\includegraphics[width=1.1in,height=1.25in,clip,keepaspectratio]{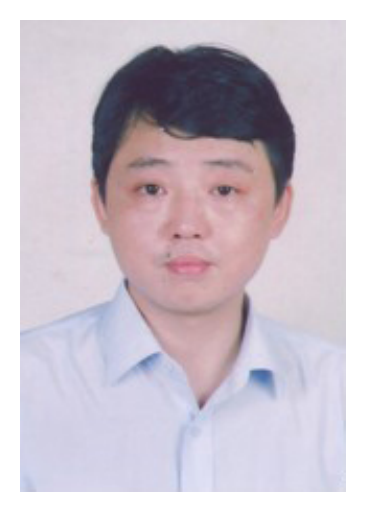}}]{Xiaopei Wu}
	is a professor at the School of Computer Science and Technology, Anhui University, and a doctoral/master's supervisor. He received his bachelor's, master's and Doctor's degrees from Anhui University, University of Electronic Science and Technology of China and University of Science and Technology of China in 1985, 1988 and 2002, respectively. 2003-2006 Postdoctoral research at the Signal and Information Processing Postdoctoral Mobile Station of the University of Science and Technology of China, 2004.4-2004.10 Study at the University of California, San Diego. Main research areas: Machine learning and brain-computer interface; Voice and intelligent video analysis; Human biological signal monitoring and special human-computer interaction technology. 
\end{IEEEbiography}

\vfill
\end{document}